\documentclass[showpacs,aps,prl,amsmath,superscriptaddress,twocolumn]{revtex4}

\usepackage{mathrsfs}
\usepackage{amssymb}
\usepackage{amstext}
\usepackage[english]{babel}
\usepackage{graphicx}
\usepackage{bm}
\usepackage{float}
\usepackage{color}

\makeatletter

\newcommand{\Rmnum}[1]{\expandafter\@slowromancap\romannumeral #1@}
\makeatother

\begin{document}

\title{Two-Component Structure in the Entanglement Spectrum of Highly
  Excited States}

\author{Zhi-Cheng Yang}

\affiliation{Physics Department, Boston University, Boston,
  Massachusetts 02215, USA}

\author{Claudio Chamon}

\affiliation{Physics Department, Boston University, Boston,
  Massachusetts 02215, USA}

\author{Alioscia Hamma}

\affiliation{Center for Quantum Information, Institute for
  Interdisciplinary Information Sciences, Tsinghua University, Beijing
  100084, People's Republic of China}

\author{Eduardo R. Mucciolo}

\affiliation{Department of Physics, University of Central Florida,
  Orlando, Florida 32816, USA}

\date{\today}

\begin{abstract}
We study the entanglement spectrum of highly excited eigenstates of
two known models that exhibit a many-body localization transition,
namely the one-dimensional random-field Heisenberg model and the
quantum random energy model. Our results indicate that the
entanglement spectrum shows a ``two-component'' structure: a universal
part that is associated with random matrix theory, and a nonuniversal
part that is model dependent. The nonuniversal part manifests the
deviation of the highly excited eigenstate from a true random state
even in the thermalized phase where the eigenstate thermalization
hypothesis holds. The fraction of the spectrum containing the
universal part decreases as one approaches the critical point and
vanishes in the localized phase in the thermodynamic limit. We use the
universal part fraction to construct an order parameter for measuring
the degree of randomness of a generic highly excited state, which is
also a promising candidate for studying the many-body localization
transition. Two toy models based on Rokhsar-Kivelson type
wave functions are constructed and their entanglement spectra are shown
to exhibit the same structure.
\end{abstract}


\maketitle


\textit{Introduction.}--Quantum entanglement, a topic of much
importance in quantum information theory, has also gained relevance in
quantum many-body physics in the past few years
\cite{AmicoRMP,EisertRMP}. In particular, the entanglement entropy
provides a wealth of information about physical states, including
novel ways to classify states of matter that do not have a local
order parameter~\cite{wen}. However, it has been realized only
recently in various physical contexts that the entanglement entropy is
not enough to fully characterize a generic quantum state. For example,
the quantum complexity corresponding to the geometric structure of
black holes cannot be fully encoded just by the entanglement entropy
\cite{Susskind}. One natural step beyond the amount of
entanglement is the specific pattern of entanglement, i.e.,
the entanglement spectrum. A recent result that motivates this
direction is the relationship between irreversibility and entanglement
spectrum statistics in quantum circuits \cite{CHM,Shaffer}. It was
shown that irreversible states display Wigner-Dyson statistics in the
level spacing of entanglement eigenvalues, while reversible states
show a deviation from Wigner-Dyson distributed entanglement levels and
can be efficiently disentangled.

Are there universal features in the entanglement spectrum of a generic
eigenstate of a quantum Hamiltonian? Highly excited eigenstates of a
generic quantum Hamiltonian are believed to satisfy the ``eigenstate
thermalization hypothesis'' (ETH) \cite{Deutsch,Srednicki,Rigol},
which states that the expectation value $\langle
\psi_{\alpha}|\hat{O}|\psi_{\alpha}\rangle$ of a few-body observable
$\hat{O}$ in an energy eigenstate $|\psi_{\alpha}\rangle$ of the
Hamiltonian with energy $E_{\alpha}$ equals the microcanonical average
at the mean energy $E_{\alpha}$. So one could as well ask the following question:  What is the
structure of the entanglement spectrum of highly excited eigenstates
of a thermalized system? Here we find a quandary. Completely random
states are generically not physical, namely, they cannot be the
eigenstates of Hamiltonians with local interactions. For the ETH to be a
physical scenario for thermalization, highly excited eigenstates of
physical local Hamiltonians cannot always be completely random, yet
they have to contain enough entropy. Deviations from a completely
random state can be quantified by the entanglement entropy, more
precisely by the amount that it deviates from the maximal entropy in
the subsystem, derived by Page, which we will refer to as the Page
entropy hereafter~\cite{Page}. But are there features that cannot be
captured by the entanglement entropy alone? Can one identify remnants
of randomness in the full entanglement spectrum? What about in states
that violate the ETH?

In this Letter, we address the above questions using as a case study the
problem of many-body localization (MBL) \cite{Nandkishore,
  oganesyan07, Pal_Huse, Moore, Abanin}. We study two known models
that were shown to exhibit a MBL transition, namely, the Heisenberg
spin model with random fields, and the quantum random energy model
(QREM) \cite{Kurchan, QREM, QREM2}. In the delocalized phase,
high-energy eigenstates are thermalized according to the ETH. The
deviation from completely random states manifests itself in a
``two-component'' structure in the entanglement spectrum: a universal
part that corresponds to random matrix theory\cite{Mehta}, and
a nonuniversal part that is model dependent. We show that the
universal part fraction decreases as one approaches the transition
point and vanishes in the localized phase in the thermodynamic
limit. We therefore propose an order parameter that is able to measure
the degree of randomness of a generic highly excited state and capture
the many-body localization-delocalization transition based on the
entanglement spectrum, and show that it gives predictions consistent
with previous results. We further construct two toy models in terms of
Rokhsar-Kivelson- (RK) type wavefunctions \cite{RK, RK_2} and the same
structure in the entanglement spectra is observed.


\textit{Heisenberg spin chain.}--A well-studied model that shows a MBL
transition is the isotropic Heisenberg spin-1/2 chain with random
fields along a fixed direction,
\begin{equation}
\mathcal{H} = \sum_{i=1}^{L} \left( h_i \,S_i^z + J\,\vec{S}_i \cdot
  \vec{S}_{i+1} +\Gamma S_i^x\right),
\label{eq1}
\end{equation}
where the random fields $h_i$ are independent random variables at each
site, drawn from a uniform distribution in the interval
$[-h,h]$. $\Gamma$ is a uniform transverse field along the
$x$ direction, which breaks total $S_z$ conservation. We assume
periodic boundary condition and set the coupling $J=1$ and
$\Gamma=0.1$. In the absence of the transverse field $\Gamma$,
previous work located the critical point at $h=h_c \approx 3.5$ in the
$S_z=0$ sector \cite{Pal_Huse, Heisenberg, Serbyn}. We consider two
different regimes by varying the disorder strength parameter $h$: (i)
within the thermalized phase ($h<h_c$), and (ii) in the localized
phase ($h>h_c$). In each regime, we focus on eigenstates of
Hamiltonian (\ref{eq1}) at the middle of the spectrum,
namely, on highly excited states.


\begin{figure*}[!ht]
\centering
\scalebox{.45}{\includegraphics[width=2.\textwidth]{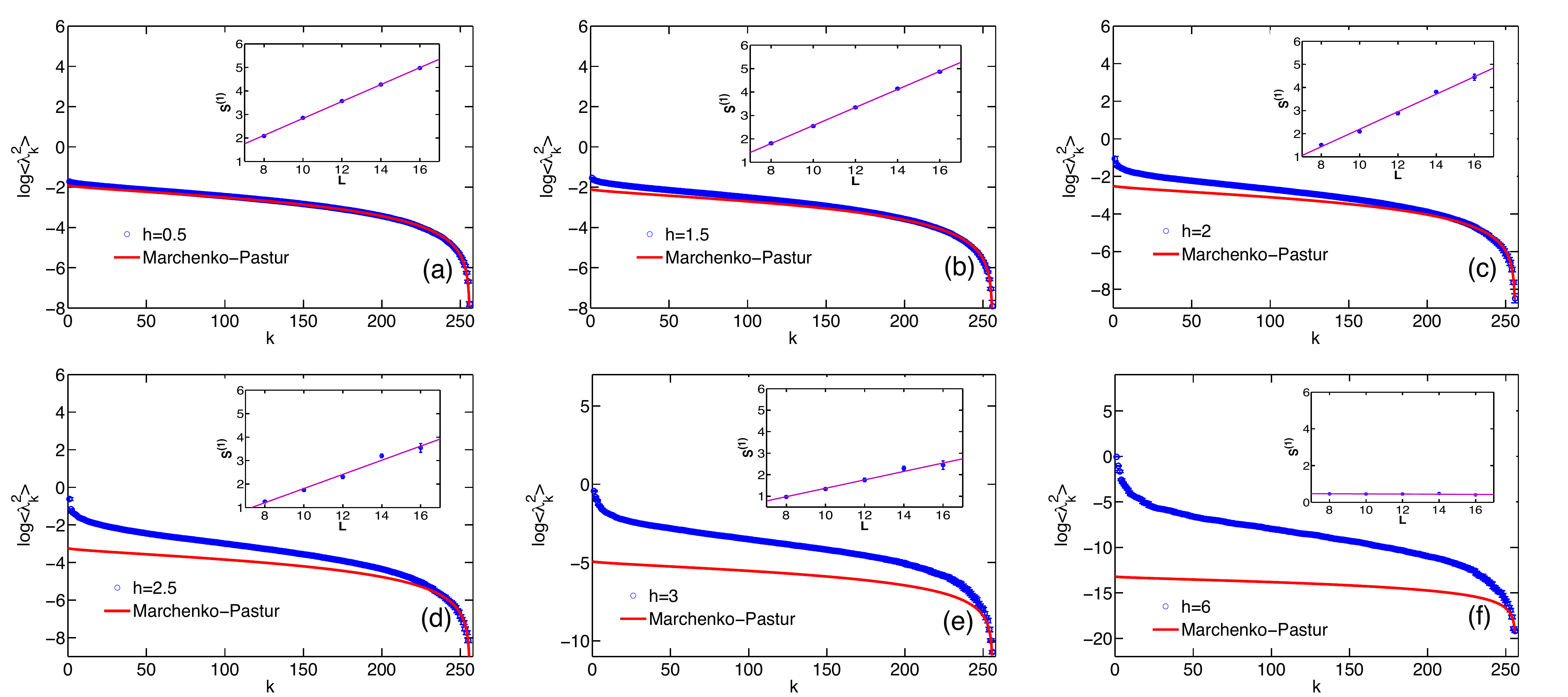}}
\caption{(Color online) Average entanglement spectrum of highly
  excited eigenstates for a system of size $L=16$, averaged over 10
  realizations of disorder (plotted in logarithmic scale). Panels a--f
  show the spectrum for $h=0.5,1.5,2,2.5,3$ and $6$, respectively. The
  solid lines correspond to the spectrum of a completely random state
  (derived from a Marchenko-Pastur distribution), and is shown for
  reference. Insets: scaling of the average entanglement entropy
  $S^{(1)}$ with system size.}
\label{f1}
\end{figure*}

We consider a bipartition of the system into subsystems $A$ and $B$ of
equal size ($L/2$ sites each). For a generic eigenstate $|\psi\rangle
= \sum_{\bm{\sigma}} \psi(\bm{\sigma}) |\bm{\sigma}\rangle$, where
$\bm{\sigma}\equiv \sigma_1\sigma_2\dots\sigma_L$ labels the $2^L$
possible spin configurations of the system, we cast the wave function
as $\psi(\bm{\sigma})\equiv \psi(\bm{\sigma}_A\,\bm{\sigma}_B)$, where
$\bm{\sigma}_A\equiv \sigma_1\dots\sigma_{L/2}$ and
$\bm{\sigma}_B\equiv \sigma_{L/2+1}\dots\sigma_L$. The entanglement
spectrum is obtained from the eigenvalues of the reduced density
matrices $\rho_A={\rm tr}_B |\psi\rangle \langle \psi|$ and
$\rho_B={\rm tr}_A|\psi\rangle \langle \psi|$: $\{p_k=\lambda_k^2\},
k=1, \ldots 2^{L/2}$. In this work, we are primarily concerned with
the density of states and level statistics of the $\{\lambda_k\}$ for
highly excited eigenstates for different strengths of disorder. For
each value of $h$ analyzed, the spectra were averaged over 10
realizations of disorder for $L=16$, and 100 realizations for
$L=14$. For each spectrum, the eigenstate with energy closest to zero
was obtained by a Lanczos projection \cite{anders}. This eigenstate
corresponds to a highly excited state.


\textit{Thermalized phase.}--We start by considering the weakly
disordered case, $h\ll h_c$. Only a small amount of disorder is
necessary to break the integrability of the clean
Hamiltonian. However, conservation of the total $S_z$ also plays a
crucial role in making eigenstates completely random. A small
transverse field $\Gamma$ is applied to break this conservation
without substantially altering the many-body localization
transition. In this regime, we find that the entanglement spectrum of
the highly excited state with eigenenergy near zero is close to that
of a completely random quantum state, as shown in Fig.~\ref{f1}(a) for
systems of size $L=16$ and $h=0.5$. The entanglement spectrum follows
closely a Marchenko-Pastur distribution (with proper normalization),
which describes the asymptotic average density of eigenvalues of a
Wishart matrix \cite{Marchenko-Pastur, Znidaric}. (The expression for
the entanglement spectral density for the random state is presented in
the Supplemental Material\cite{SM}.) One can also check that, in this regime,
the von Neumann entanglement entropy $S^{(1)} = -\sum_{k}p_k\ln p_k$
is in good agreement with the Page entropy for random states:
$S_{m,n}=\sum_{k=n+1}^{mn}\frac{1}{k}-\frac{m-1}{2n}\approx {\rm
  ln}(m)-\frac{m}{2n}$, where $m$ and $n$ are the Hilbert space
dimensions of subsystem $A$ and $B$, respectively \cite{Page}. For
example, our computed average entropy for 16 sites is $\langle S^{(1)}
\rangle = 4.9719 \pm 0.0015$, while the corresponding Page entropy is
$S_{\rm Page} = 5.0452$.

As the disorder strength is increased, but still $h<h_c$, the system
remains in the thermalized phase where it is supposed to obey the ETH
and yield volume-law scaling of the entanglement entropy with
system sizes~\cite{Deutsch_Li_Sharma}, which is verified in the insets
of Figs. \ref{f1}(a) to \ref{f1}(e). However, in spite of the volume-law
scaling of the entanglement entropy and the thermalization of
eigenstates, the entanglement entropy is much lower than the Page
entropy. This indicates that the pattern of entanglement must have
changed, which is manifest in the spectra shown in Figs. \ref{f1}(b) to
\ref{f1}(e). The entanglement spectrum shows a striking
``two-component'' structure: (i) a universal tail in agreement with random matrix theory, and (ii) a nonuniversal part. The non-universal part dominates
the weights in the spectrum (large $\lambda_k$ values), resulting in
low entanglement entropy, as it decays much faster than the universal
part. Therefore, we find that although thermalized states are
not necessarily random states, they partially retain a
component that is reminiscent of a random state: the entanglement
spectrum follows the Marchenko-Pastur level density distribution. In
addition, the universal part of the entanglement spectrum follows a
Wigner-Dyson distribution of level spacings (see the Supplemental
Material\cite{SM}).


\textit{Localized phase.}--In this regime, the entanglement entropy
exhibits an area-law scaling with the system size [see inset
of Fig. \ref{f1}(f)], which in one spatial dimension
implies a constant entropy and, at most, weakly logarithmic
corrections, in accordance with Ref. \cite{Bauer}.

The entanglement spectrum in the localized regime, depicted in
Fig.~\ref{f1}(f) for $h=6$, shows a different scenario from that in the
thermalized phase: the universal part of the spectrum disappears
completely, leaving only the nonuniversal part characterized by its
fast decay rate.

\textit{QREM.}--The QREM describes $L$ spins in a transverse field
$\Gamma$ with the following Hamiltonian:
\begin{equation}
\mathcal{H} = E(\{\sigma^z\})+\Gamma \sum_{i=1}^{L} \sigma_i^x
\end{equation}
where $E(\{\sigma^z\})$ is the classical REM term that takes independent values
from a Gaussian distribution of zero mean and variance
$L/2$ \cite{Derrida}. This model was first studied in the context of a
mean-field spin glass, and was shown to exhibit a first-order quantum
phase transition as a function of $\Gamma$ \cite{Kurchan}. More
recently, it was further demonstrated to have a MBL transition when
viewed as a closed quantum system \cite{QREM}. Numerical and
analytical arguments show that the transition happens at an energy
density $|\epsilon|=\Gamma$ in the microcanonical ensemble. Since
there is no support for the many-body localized phase at energy
density $\epsilon=0$, we examine the eigenstates with energy density
closest to $\epsilon=0.5$ instead, and study the entanglement spectrum
as $\Gamma$ is tuned. The two-component structure and its evolution as
a function of $\Gamma$ similar to Fig. \ref{f1} are again observed
(see the Supplemental Material\cite{SM}).


\textit{An order parameter.}--The above picture unveils a new aspect of
the MBL transition. The two parts of the entanglement spectrum of a
highly excited state clearly evolve as the disorder strength $h$ is
increased, namely, the universal part shrinks and the nonuniversal
part grows. This fact suggests that one could use the fraction of each
component as an order parameter.

Figures~\ref{f1}(a) to \ref{f1}(e) indicate an $h$ dependent value $k_h$
that separates the nonuniversal ($k\leq k_h$) from the universal ($k>
k_h$) parts of the rank-ordered entanglement levels (see the Supplemental
Material \cite{SM}  for the protocol for determining $k_h$). One can thus define
the partial R\'enyi entropies
\begin{equation}
S_{\leq}^{(q)} = \frac{1}{1-q} \ln \sum_{k\leq k_h} \; p_k^q \;,
\end{equation}
with $q\geq 0$. Because the universal part of the spectrum is where
the eigenvalues with low entanglement reside, this part of the
spectrum is obscured by any measure that relies on the eigenvalues as
weights. A good measure of the fraction of the two components that
does not depend on these weights is given by the $q=0$ R\'enyi entropy, 
which simply measures the ranks: $S_{\leq}^{(0)} = \ln
k_h$. Therefore, an order parameter that measures the fraction of the
universal component is
\begin{equation}
{\cal O}_{\rm MBL} = 1-\frac{S_{\leq}^{(0)}}{S^{(0)}} = 1 - \frac{\log_2
  k_h}{{L/2}} \;.
\end{equation}
%

\begin{figure}[!ht]
\centering
\includegraphics[width=0.37\textwidth]{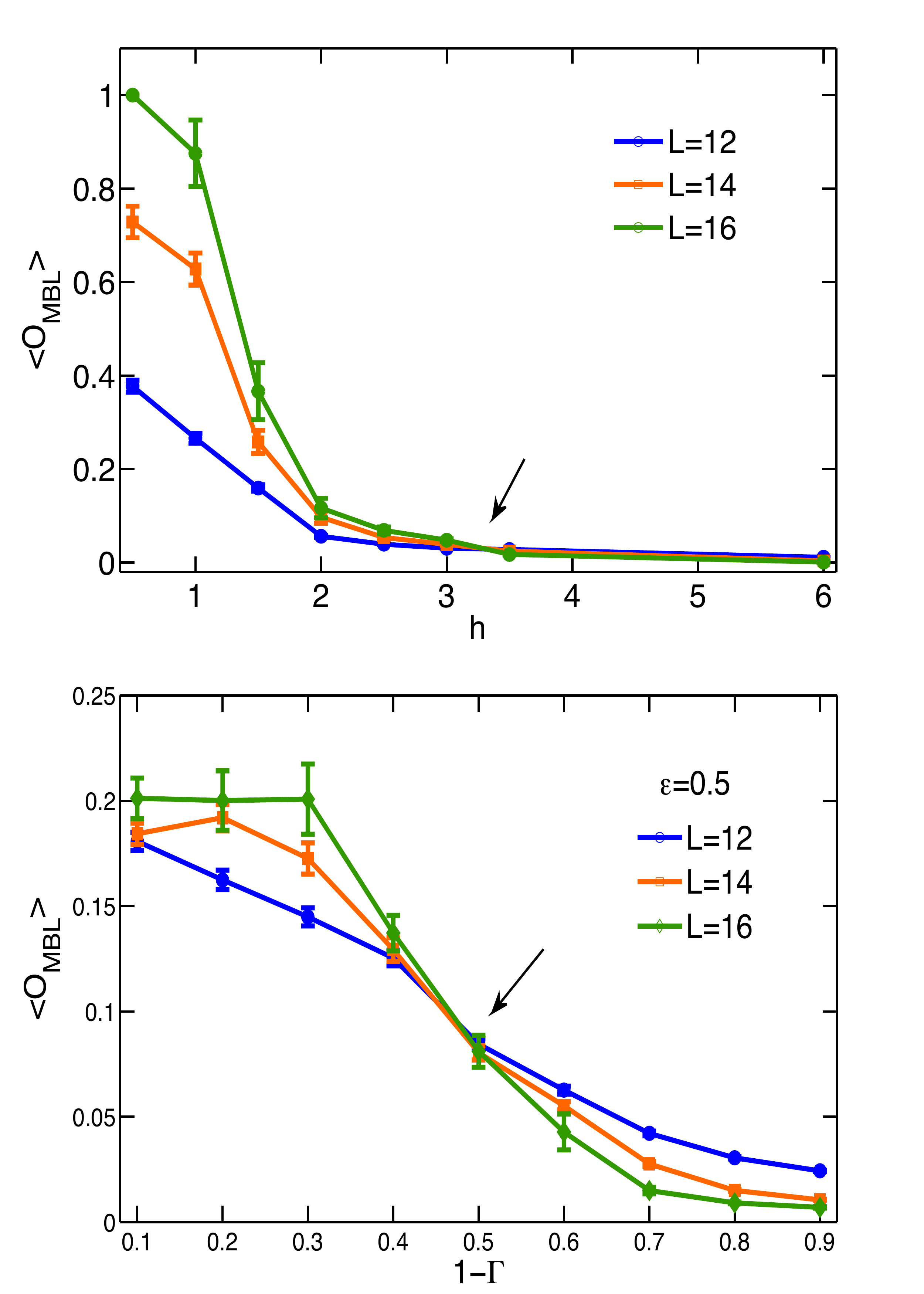}
\caption{(Color online) The order parameter defined as the fraction of
  the universal component in the full entanglement spectrum for the
  Heisenberg spin model (upper panel) and the QREM (lower panel).}
\label{fig:orderparameter}
\end{figure}

Figure \ref{fig:orderparameter} shows the order parameter as defined
above for the Heisenberg spin model and the QREM, respectively. For
the QREM, all curves at different system sizes cross at $\Gamma_c
\approx 0.5$, in excellent agreement with Ref. \cite{QREM}. We have
also looked at energy density $\epsilon=0.3$, and the curves cross at
$\Gamma_c\approx 0.25$, giving the same numerical prediction as in
Refs. \cite{QREM} and \cite{QREM2} (plot shown in the Supplemental
Material \cite{SM}). For the random-field Heisenberg model, however, the fact
that the transition happens at the point where the order parameter is
nearly zero makes it harder to accurately locate the critical point
using our order parameter. We see from Fig. \ref{fig:orderparameter}
that the curves cross at $h_c \approx 3.3$, which is also consistent
with previous studies. This indicates that, by considering the full
entanglement spectrum at high energies, our order parameter reveals a
novel property that is promising for studying the MBL transition.

We remark that, although the MBL transition can also be captured by
the scaling property of the entanglement entropy, our order parameter
seems to be applicable even for models with nonlocal interactions,
which could obscure the connection between the volume-to-area law
transition of the entropy and the MBL transition.


\textit{Toy models.}--we construct two RK-type model wave functions
that are shown to have (i) the two-component structure in their
entanglement spectra, and (ii) a phase transition as a function of the
tuning parameter. The wave functions take the following form:
\begin{equation}
|\Psi\rangle = \frac{1}{\sqrt{\mathcal{Z}}} \sum_{{\bm \sigma}}
s_{{\bm \sigma}} {\rm e}^{-\frac{\beta}{2} E({\bm \sigma})} |{\bm
  \sigma}\rangle,
\end{equation}
where $E({\bm \sigma})$ is the energy for the classical configuration
${\bm \sigma}$ and $\mathcal{Z}$ is the corresponding partition
function of the classical statistical system \cite{RK_2}. $s_{\bm
  \sigma}$ is a random sign for each configuration, such that the
wave function represents a highly excited state. We consider the
following two cases: (i) $E({\bm \sigma})=E_{{\rm REM}}({\bm
  \sigma})$, and (ii) $E({\bm \sigma})=-\frac{J}{L}\sum_{i<j} \sigma_i^z
\sigma_j^z$. In the first case, the energy is taken to be that of the
REM, while in the second case the energy is that of an infinite-range
uniform ferromagnetic interaction.

\begin{figure}[!hb]
\centering
\includegraphics[width=0.37\textwidth]{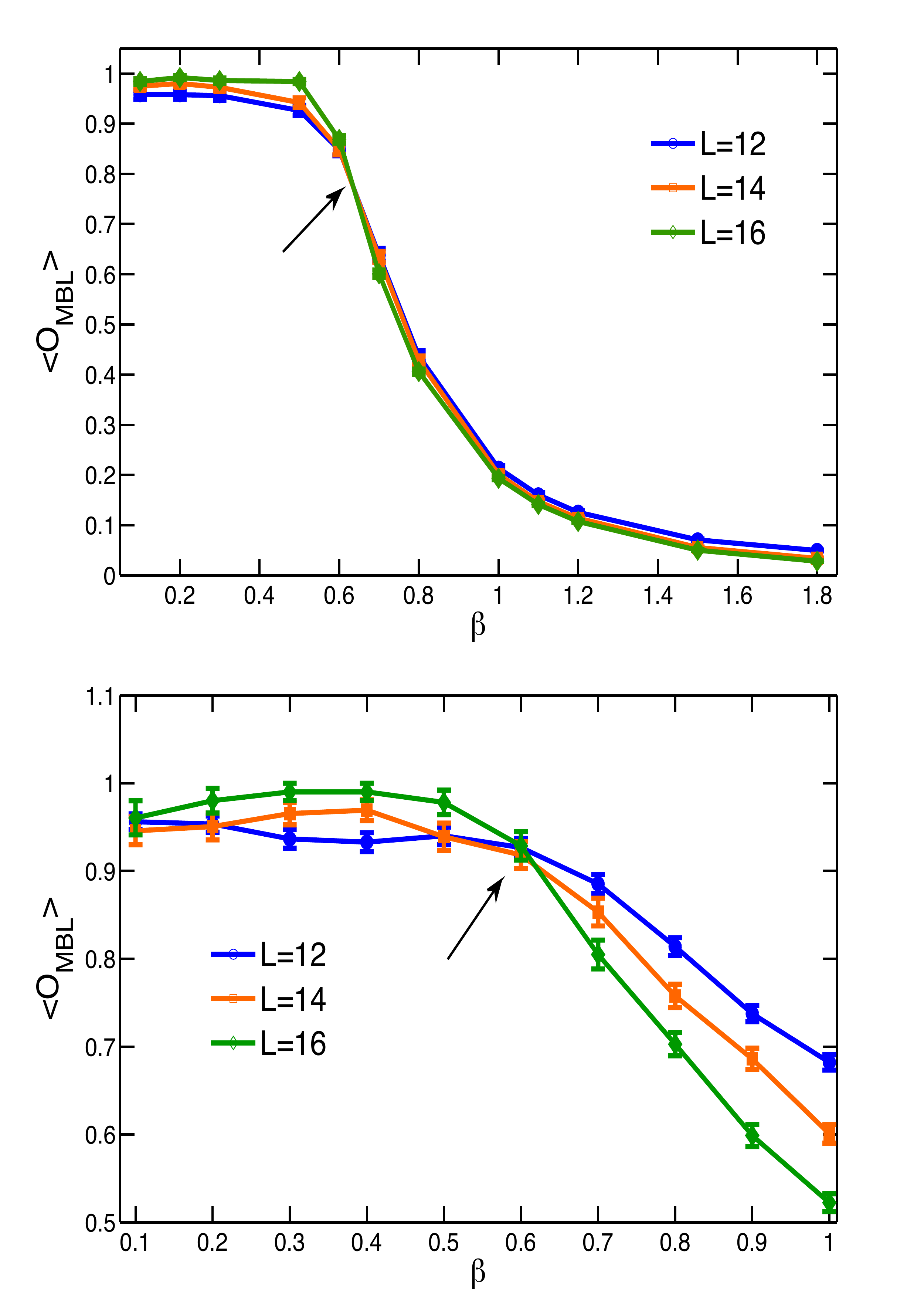}
\caption{(Color online) The order parameter for the random-sign
  RK-type wave functions. Upper panel: $E({\bm \sigma})=E_{{\rm
      REM}}({\bm \sigma})$. Lower panel: $E({\bm
    \sigma})=-\frac{J}{L}\sum_{i<j} \sigma_i^z \sigma_j^z$, with $J=1$.}
\label{f3}
\end{figure}

In the small $\beta$ regime, the above RK-type wave functions are close
to completely random states; upon increasing $\beta$, the
wave functions are pushed towards product states and start to deviate
from completely random states. Therefore, the tuning parameter $\beta$
here plays the role of the ``disorder strength''. Indeed, we find the
same two-component structure in the entanglement spectrum (see the 
Supplemental Material \cite{SM}), and the order parameter is shown in
Fig. \ref{f3}. The REM case was recently studied by Chen \textit{et
  al.} where the MBL transition was obtained numerically using other
measures \cite{Fradkin}. Here we clearly see that, in both cases, the
curves cross at some critical $\beta$, indicating the existence of a similar phase transition.


\textit{Summary and discussion.}--The details of the structure of the
entanglement spectrum, especially the universal part at the tails of
the spectrum, have long been overlooked. The main focus has been
primarily on the dominating nonuniversal component, and the universal
tail has thus far been discarded. For example, in the density matrix
renormalization group \cite{DMRG} and tensor network methods
\cite{MPS}, the density matrix is truncated to avoid uncontrolled
growth of its dimensions. While this procedure is certainly justified
when the purpose is to obtain ground state properties, it discards
important information about the behavior of the system at higher
energy states. In this Letter we showed that the full entanglement
spectrum, directly computable from the wave function, provides
information that is often invisible in the entanglement entropy
alone.

On the other hand, much has been known about random quantum states,
e.g., the Page entropy and volume-law scaling entropy. Nevertheless,
the Page entropy is often an overestimate of the actual entanglement
entropy computed from generic quantum states. Therefore, a natural
question that arises is as follows: How random does a given quantum state look?
In this Letter, we show that a generic quantum state that satisfies
ETH does not necessarily mean a completely random state. We present an
order parameter to quantify the degree of randomness by using
information about the full entanglement spectrum. In the context of
MBL, our order parameter is able to locate the critical point,
consistent with previous results. Our work may provide a novel way of
studying MBL, and may shed new light on the understanding many-body
systems at the level of wave functions.


Z.-C.Y. is indebted to Bernardo Zubillaga, Alexandre Day, Shenxiu Liu,
and Yi-Zhuang You for generous help and useful discussions. We thank
Christopher Laumann for useful comments. This work was supported in
part by DOE Grant No. DEF-06ER46316 (C.C.), by NSF Grant No. CCF 1117241
(E. R. M.), by National Basic Research Program of China Grants No.
2011CBA00300 and No. 2011CBA00301, and by National Natural Science Foundation
of China Grant No. 61361136003 (A.H.)



\clearpage
\onecolumngrid

\section{Supplemental Material}

\section{Marchenko-Pastur distribution and random states}

The Marchenko-Pastur distribution describes the asymptotic (large-$N$)
average density of eigenvalues of an $N\times N$ matrix of the form
$Y=XX^\dagger$, known as a Wishart matrix, where $X$ is a $N\times M$
random rectangular matrix with independent but identically distributed
entries \cite{Marchenko-Pastur}. Let $\sigma^2$ be the variance of the
entries in $X$. When $N=M\rightarrow \infty$, the Marchenko-Pastur
distribution takes the form
\begin{equation}
{\cal D}(p) = \left\langle \frac{1}{N} \sum_{k=1}^N \delta(p - p_k)
\right\rangle_{N\rightarrow\infty} = \frac {2}{\pi p_{\rm max}}
\sqrt{\frac{p_{\rm max}}{p} -1},
\end{equation}
where $\{p_k\}$ are the eigenvalues of $Y$, $0\leq p \leq p_{\rm max}$
and $p_{\rm max} = 2\sigma^2$. From this distribution we can obtain
the average number function associated to the eigenvalues of $Y$. Let
$p_1\geq p_2\geq \cdots \geq p_N$ and $\eta_k = k/N$. Then,
\begin{eqnarray}
\eta(p) & = & 1 - \left\langle \frac{1}{N} \sum_{k=1}^N \theta(p-p_k)
\right\rangle_{N\rightarrow\infty} \\ & = & \left[ 1 - \int_0^p
  dp^\prime\, {\cal D}(p^\prime) \right] \\ & = & \int_p^{p_{\rm max}}
dp^\prime\, {\cal D}(p^\prime) \\ & = & 1 - \frac{2}{\pi} \left[ u
  \sqrt{1-u^2} + \mbox{arcsin}(u) \right]_{u=\sqrt{p/p_{\rm max}}}.
\end{eqnarray}
Thus, introducting the rescaled variable $x=\sqrt{p/p_{\rm max}}$, we find
\begin{equation}
\label{eq:numberfunction}
\eta(x) = 1 - \frac{2}{\pi} \left[ x \sqrt{1-x^2} + \mbox{arcsin}(x)
  \right].
\end{equation}

It is straightforward to relate the average number function derived
from the Marchenko-Pastur distribution with that obtained from the
entanglement spectrum of a bipartitioned random vector. Let
$\psi(x_A,x_B)$ be the wavefunction of the bipartite system. Then, the
reduced density matrix is given by
\begin{equation}
\rho_A(x_A,x_A') = \sum_{x_B} \psi(x_A,x_B)\, \psi^\ast(x_A',x_B).
\end{equation}
We can see that, for completely random wavefunctions, the reduced
density matrix is a random Wishart matrix and therefore its
eigenvalues should follow a Marchenko-Pastur
distribution\cite{Znidaric}. Thus, we expect the average number
function to provide an accurate description of the average spectrum.

Let $\{p_k\}$, $k=1,\ldots,d$, be the set of eigenvalues of $\rho_A$
in decreasing order, with $p_k\geq 0$, $\sum_{k=1}^d p_k = 1$, and
$d\leq 2^{L/2}$. It is straightforward to relate the eigenvalues
$\{p_k\}$ to the singular values $\{\lambda_k\}$ resulting from the
Schmidt decomposition of the bipartite wavefunction,
\begin{equation}
\psi(x_A,x_B) = \sum_{k=1}^d \lambda_k\, \phi_A^{(k)}(x_A)\,
\phi_B^{(k)}(x_B),
\end{equation}
by simply setting $p_k = \lambda_k^2$ (notice that $\lambda_k\geq 0$),
where $\phi_A^{(k)}(x_A)$ and $\phi_B^{(k)}(x_B)$ are the
left-singular and right-singular vectors, respectively. For the
purpose of comparing the average spectra to
Eq. (\ref{eq:numberfunction}), it is necessary to rescale the singular
values and their indices as follows:
\begin{equation}
p_k = \frac{4}{d}x^2(\eta_k)
\end{equation}
where $x(\eta)$ is the inverse function of $\eta(x)$. The prefactor is
chosen to guarantee the normalization of $p_k$:
\begin{eqnarray}
\sum_{k=1}^d p_k & \rightarrow & 4 \int_{0}^1 {\rm d}\eta\, x^2(\eta)
\nonumber \\ & = & \frac{16}{\pi} \int_0^1 {\rm d}x\, x^2 \sqrt{1-x^2}
\nonumber \\ & = & 1.
\end{eqnarray}

\begin{figure*}[!hb]
\centering 
\includegraphics[width=0.7\textwidth]{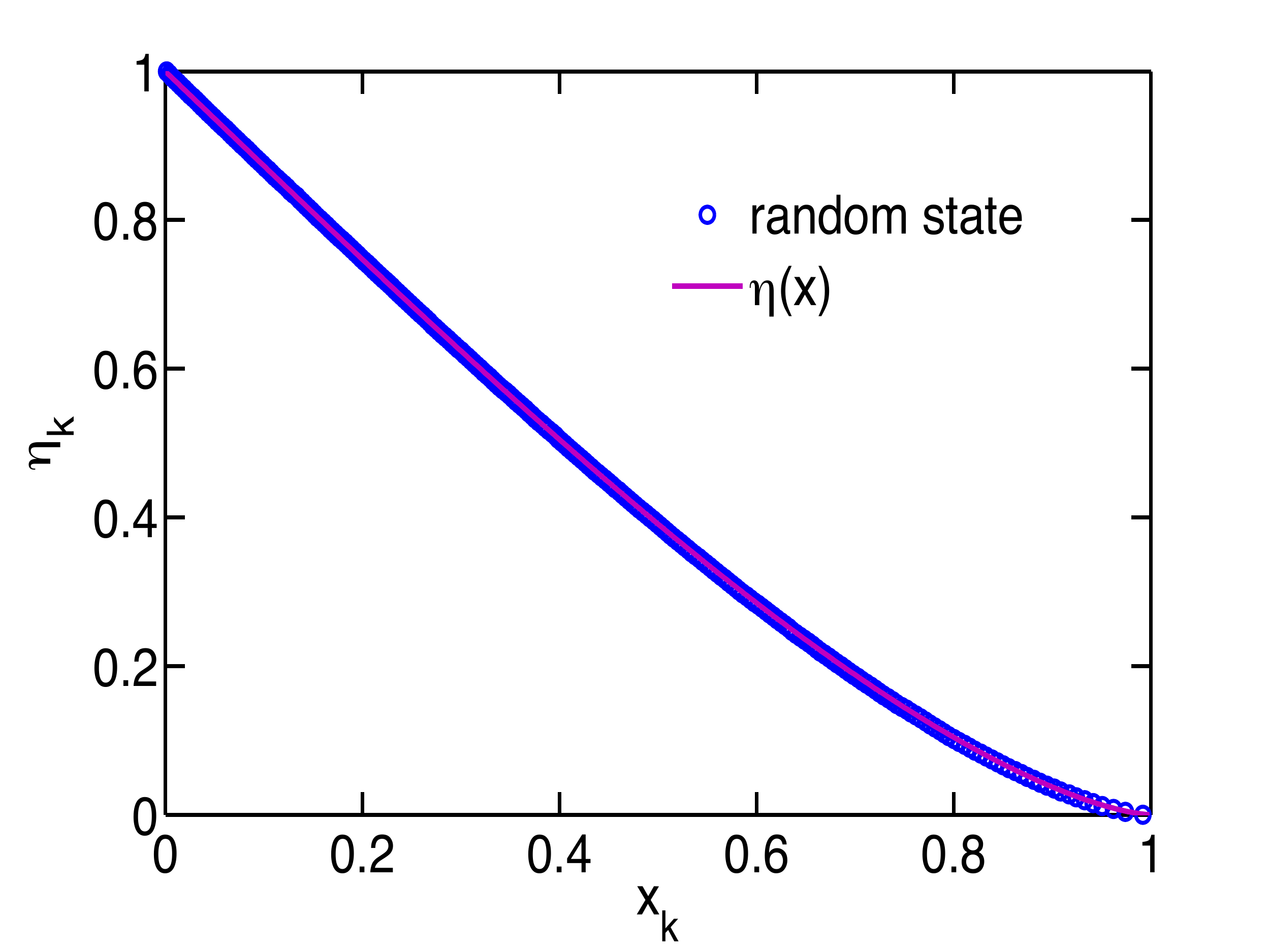}
\caption{Average entanglement pectrum of completely random wavefunction
  ($L=16$, 100 realizations used).}
\label{fig:1}
\end{figure*}

We tested this formulation by plotting the numerical results for $p_k$
obtained from a random state against the analytical expression in
Eq. (\ref{eq:numberfunction}). Figure \ref{fig:1} shows $\eta_k =
\frac{k}{d}$ versus $x_k = \frac{1}{2} \sqrt{p_kd}$. There is very
good agreement with the analytical prediction.

Notice that in the cases considered in the main text, the random part
of the entanglement spectrum alone is not normalized. However, by
plotting ${\rm log}\lambda_k$ versus $k$, the missing normalization
factor only amounts to a trivial shift of the entire spectrum.

\section{Level spacing statistics}

We studied the statistics of the entanglement spectrum by looking at
the level spacing distribution in the set $\{\lambda_k\}$. To avoid
having to perform spectral unfolding, we chose to evaluate the
distribution of ratios of adjacent level spacings \cite{oganesyan07}:
$r_k = (\lambda_{k+1} - \lambda_k)/(\lambda_k -
\lambda_{k-1})$. Accurate surmises exist for the distribution of these
ratios in the case of Gaussian ensembles \cite{atas13}. They are given
by
\begin{equation}
P_{\rm WD}(r) = \frac{1}{Z}
\frac{(r+r^2)^{\beta}}{(1+r+r^2)^{1+3\beta/2}},
\end{equation}
where $Z=\frac{8}{27}$ for the Gaussian Orthogonal Ensemble (GOE) with
$\beta=1$, and $Z=\frac{4}{81}\frac{\pi}{\sqrt{3}}$ for the Gaussian
Unitary Ensemble (GUE) with $\beta=2$. The corresponding distribution
for the Poisson distributed spectrum is given by the exact form
\begin{equation}
P_{\rm Poisson} = \frac{1}{(1+r)^2}.
\end{equation}
Notice that for the Gaussian ensembles, level repulsion manifests
itself in the asymptotic behavior $P(r\rightarrow 0) \sim r^\beta$,
which is absent in the case of Poisson statistics.

\begin{figure*}[h]
\centering
\includegraphics[width=1.\textwidth]{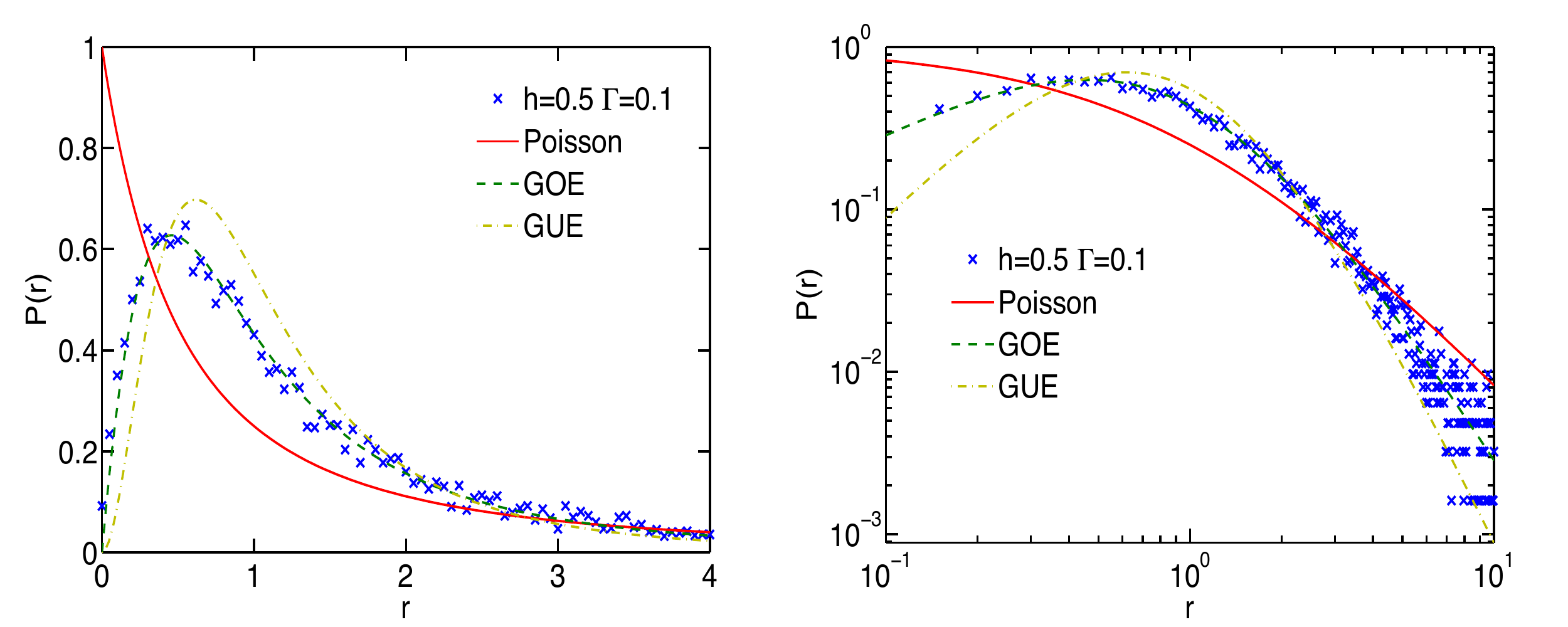}
\caption{Left panel: Distribution of the ratios of consecutive
  spacings for the entanglement spectrum of Heisenberg spin chains
  with disorder parameter $h=0.5$ and transverse field $\Gamma=0.1$
  (crosses). Right panel: the same data in a logarithmic scale. $L=16$
  and 500 realizations used.}
\label{fig:2}
\end{figure*}

Results are shown in Fig. \ref{fig:2} for disordered Heisenberg chains
with $L=16$ and 100 disorder realizations. A completely random real
state follows a GOE statistics. The universal part of the spectrum at
$h=0.5$ also follows a GOE distribution.

\section{Protocol for determining $k_h$}

The definition of the order parameter in this Letter required a
protocol for determining the point $k_h$ which separates the
non-universal part ($k \leq k_h$) from the universal ($k>k_h$) part of
the rank-ordered entanglement levels. We use the following protocol:
we took the spectrum obtained from each random state considered and
multiply it by a factor $s$ such that the rescaled smallest singular
value coincided in value with that obtained from a completely random
state. Then we swept through the spectrum, starting from the tail, and
computed the relative deviation from the completely random state
prediction, until it exceeded a certain amount. That is, until
\begin{equation}
\frac{(\lambda_k)^2-s(\lambda_k^{MP})^2}{s(\lambda_k^{MP})^2} > \epsilon,
\end{equation}
where $\epsilon$ is a number of order 1. In our case, we set $\epsilon
=1$.

\begin{figure*}[!ht]
\centering
\includegraphics[width=1\textwidth]{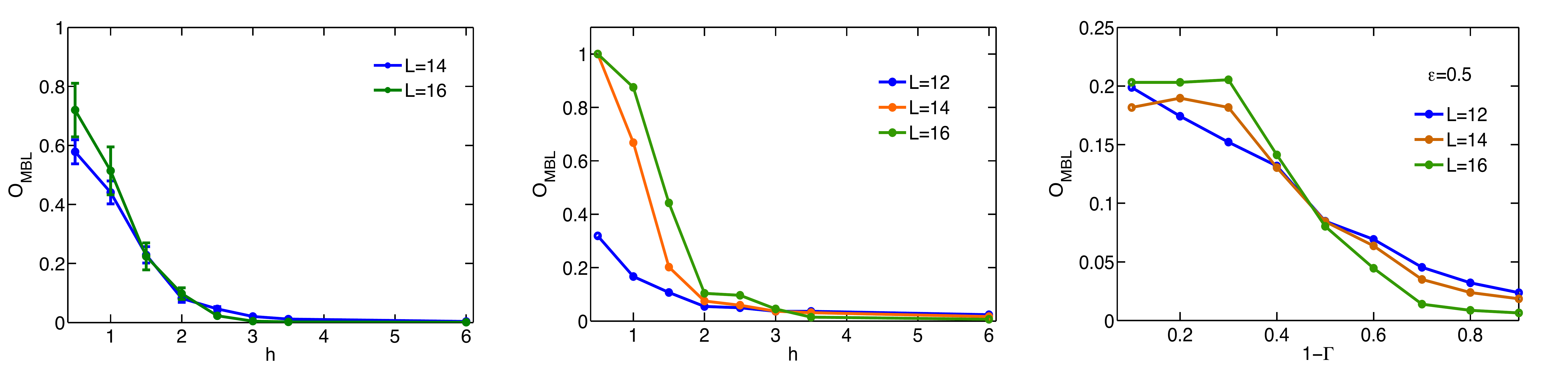}
\caption{Left: Results obtained when the matching point is chosen too
  close to the last eigenvalue for each entanglement spectrum,
  yielding a wrong transition point under finite-size
  scaling. Middle: the order parameter computed from the {\it
    averaged} entanglement spectrum of the random-field Heisenberg
  model, where fluctuations are smoothened out and not many small
  eigenvalues need to be eliminated. Right: the order parameter
  computed from the average entanglement spectrum of the QREM. One can
  see that in this case our approach is robust and the result is very
  close to that shown in the main text.}
\label{add}
\end{figure*}

However, we would like to point out one subtlety of this
methodology. In cases where the universal component of the spectrum
almost vanishes near the transition, it is hard to accurately locate
the critical point. That is because the last few points at the tail of
the spectrum show large sample-to-sample fluctuations, and our
protocol requires strictly matching a single point close to the
tail. Therefore, $k_h$ can be very sensitive to our choice of the
matching point and can even yield incorrect predictions under
finite-size scaling. On the other hand, we find that in cases such as
the QREM, where there is still a large fraction of the universal
component at the transition, this methodology is not very sensitive to
the choice of matching point. For the Heisenberg model, we locate the
critical point $k_h$ by choosing the matching point away from the tail
end, thus effectively discarding the smallest eigenvalues. For
example, for $L=16$ we discarded the last 16 eigenvalues. In order to
demonstrate that this does not lead to a sizable errors in determining
the MBL critical point, we also computed the the order parameter
directly from the \textit{averaged} entanglement spectrum, where the
fluctuations are smoothened out (Fig. \ref{add}). The critical point
found this way is very close to the one we show in the main text. We
believe that the result can be further improved if more realizations
are include in the averaring, which we hope to attempt in the future.

\section{Entanglement spectra for QREM and RK-type toy models}

\begin{figure*}[!h]
\centering
\includegraphics[width=0.8\textwidth]{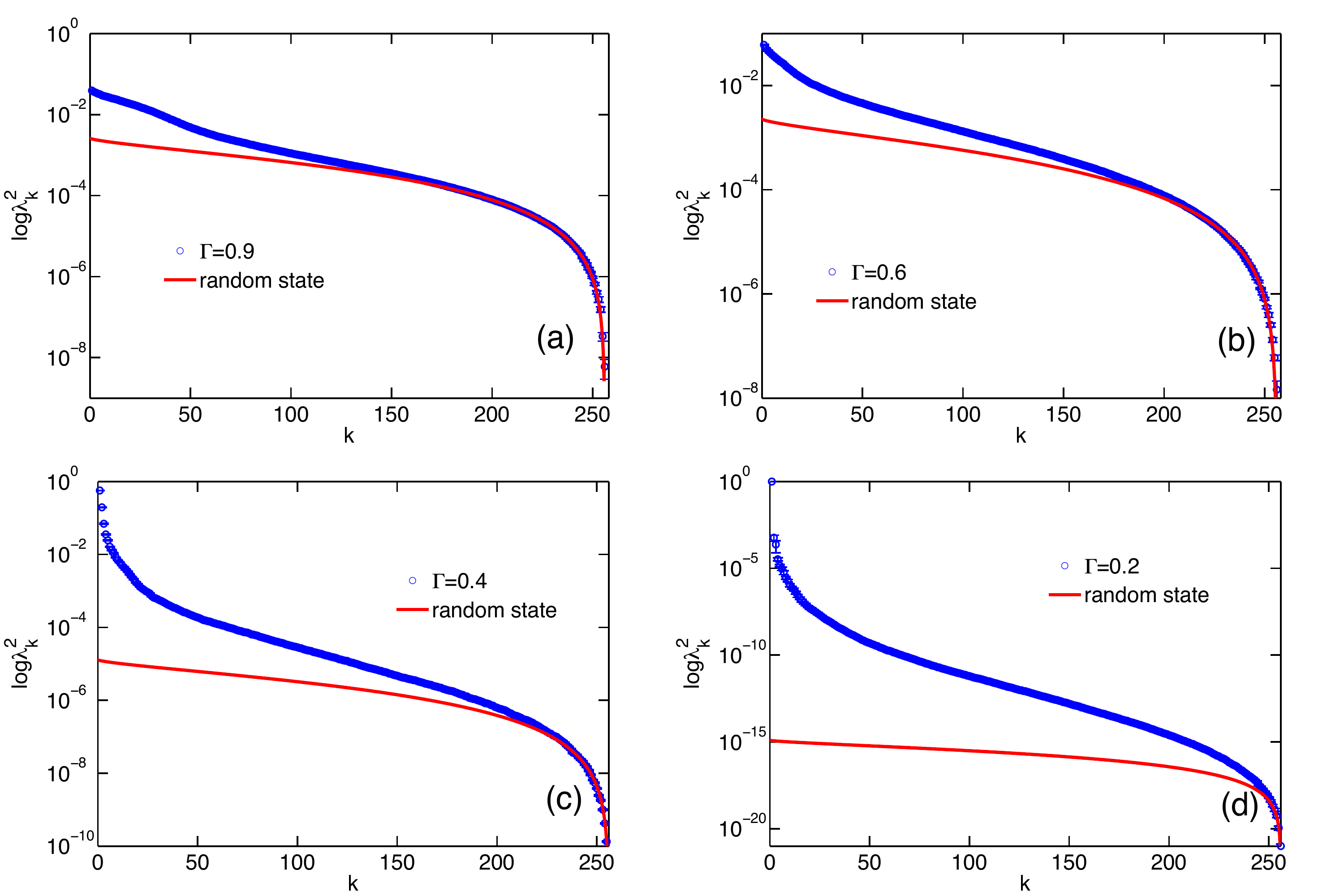}
\caption{Averaged entanglement spectrum of eigenstates with energy
  density $\epsilon=0.5$, for the QREM. The system size $L=16$, and
  $\Gamma=0.9,\ 0.6,\ 0.4$, and 0.2, averaging over 10 realizations of
  disorder.}
\label{fig:3}
\end{figure*}

\begin{figure*}[!h]
\centering
\includegraphics[width=0.8\textwidth]{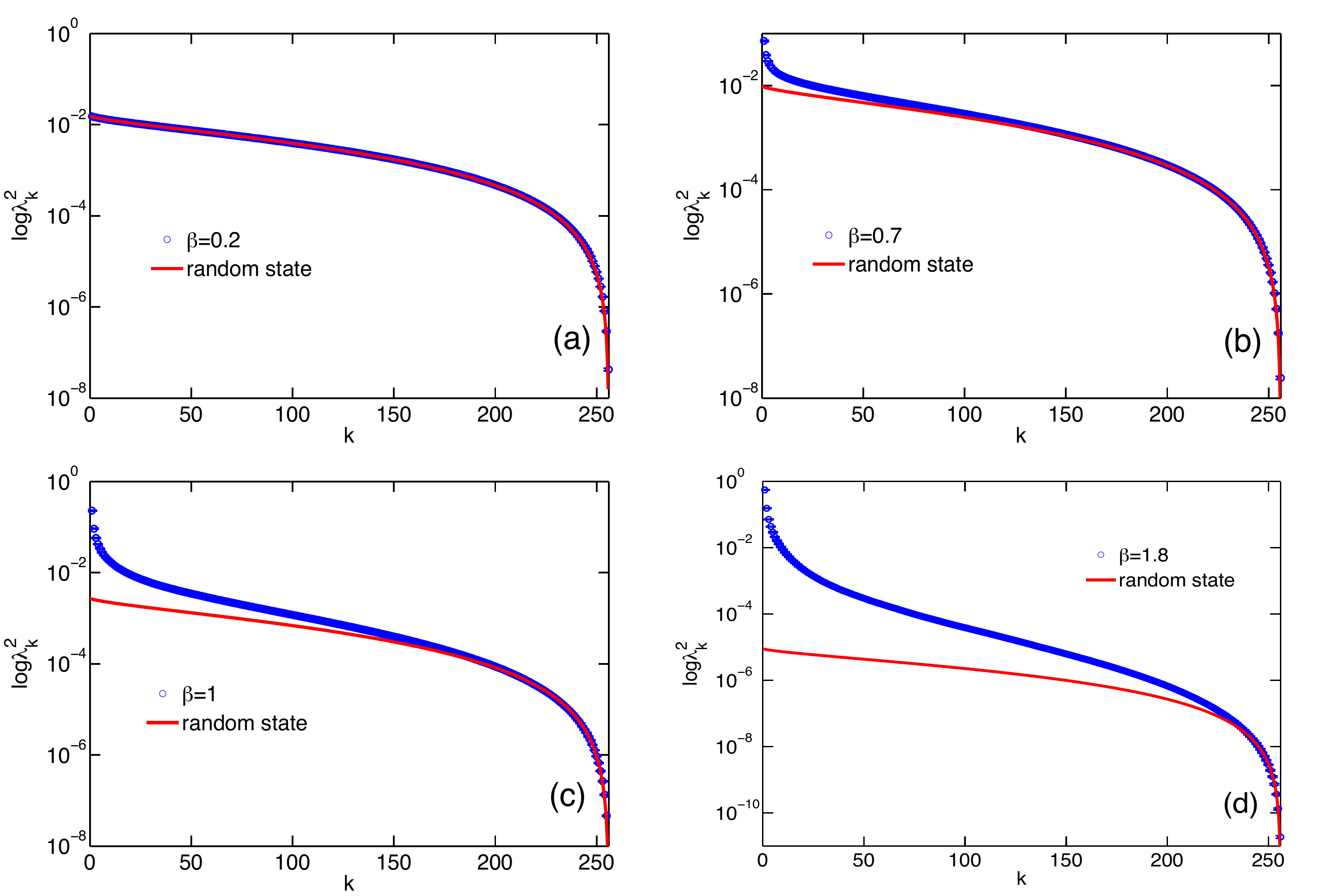}
\caption{Averaged entanglement spectrum of the RK-type wavefunction,
  with $E({\bm \sigma})=E_{{\rm REM}}({\bm \sigma})$. The system size
  $L=16$, and $\beta=0.2,\ 0.7, \ 1$, and 1.8, averaging over 100
  realizations of disorder.}
\label{fig:4}
\end{figure*}

\begin{figure*}[!h]
\centering
\includegraphics[width=0.8\textwidth]{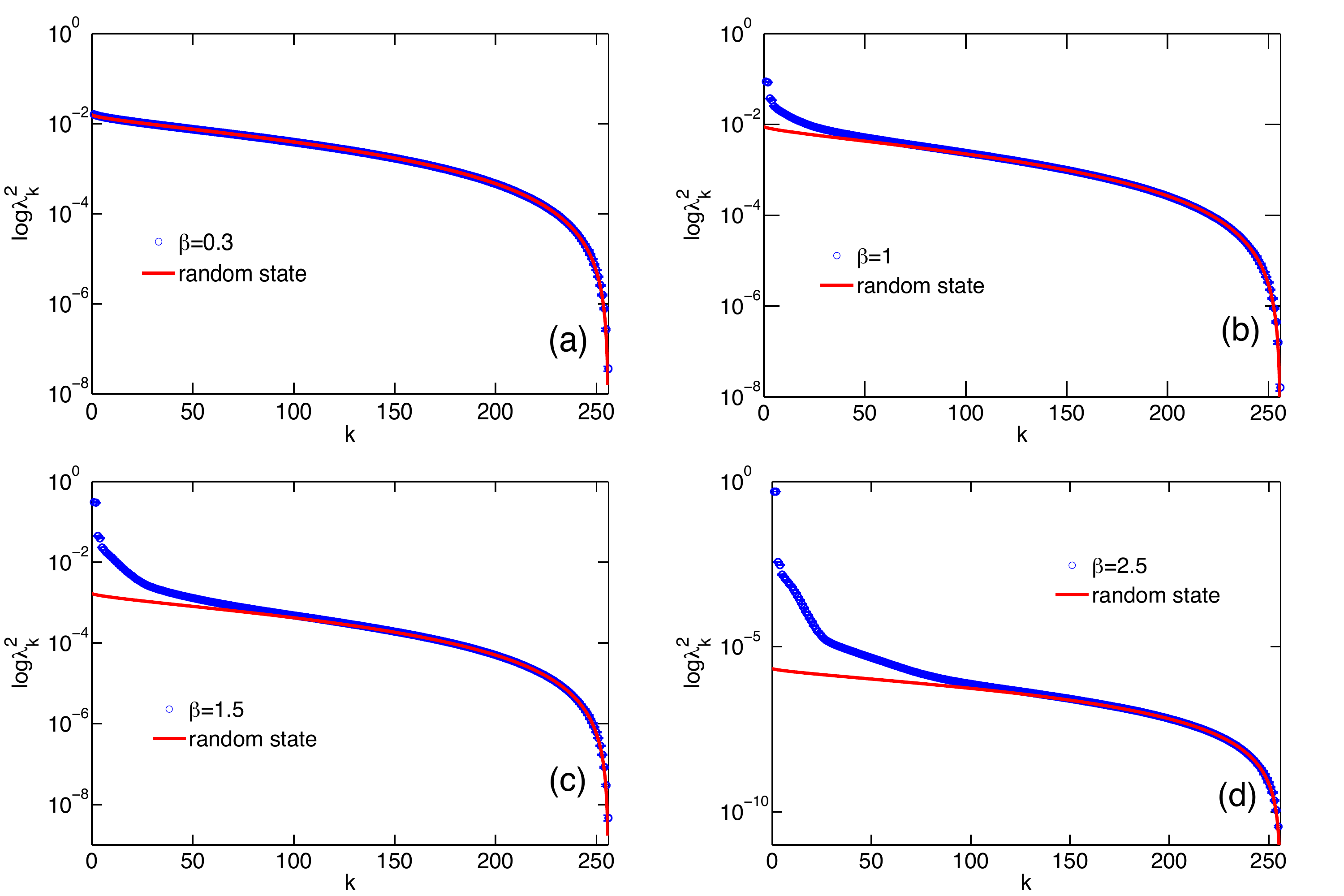}
\caption{Averaged entanglement spectrum of the RK-type wavefunction,
  with $E({\bm \sigma})=-\frac{J}{L}\sum_{i<j} \sigma_i^z
  \sigma_j^z$. The system size $L=16$, and $\beta=0.3,\ 1, \ 1.5$, and
  2.5, averaging over 100 realizations of random sign.}
\label{fig:5}
\end{figure*}

In this section, we present in Fig. \ref{fig:3} through
Fig. \ref{fig:5} the entanglement spectra for the QREM and two RK-type
toy models that were discussed in the main text.

We clearly see the following: (1) the two-component structure shows up
in all three cases; (2) the same evolution behavior as explained in
the main text happens here as well. Namely, the universal fraction
shrinks as one increases the strength of disorder, thereby pushing the
states further away from completely random states.

\begin{figure*}[!h]
\centering
\includegraphics[width=0.5\textwidth]{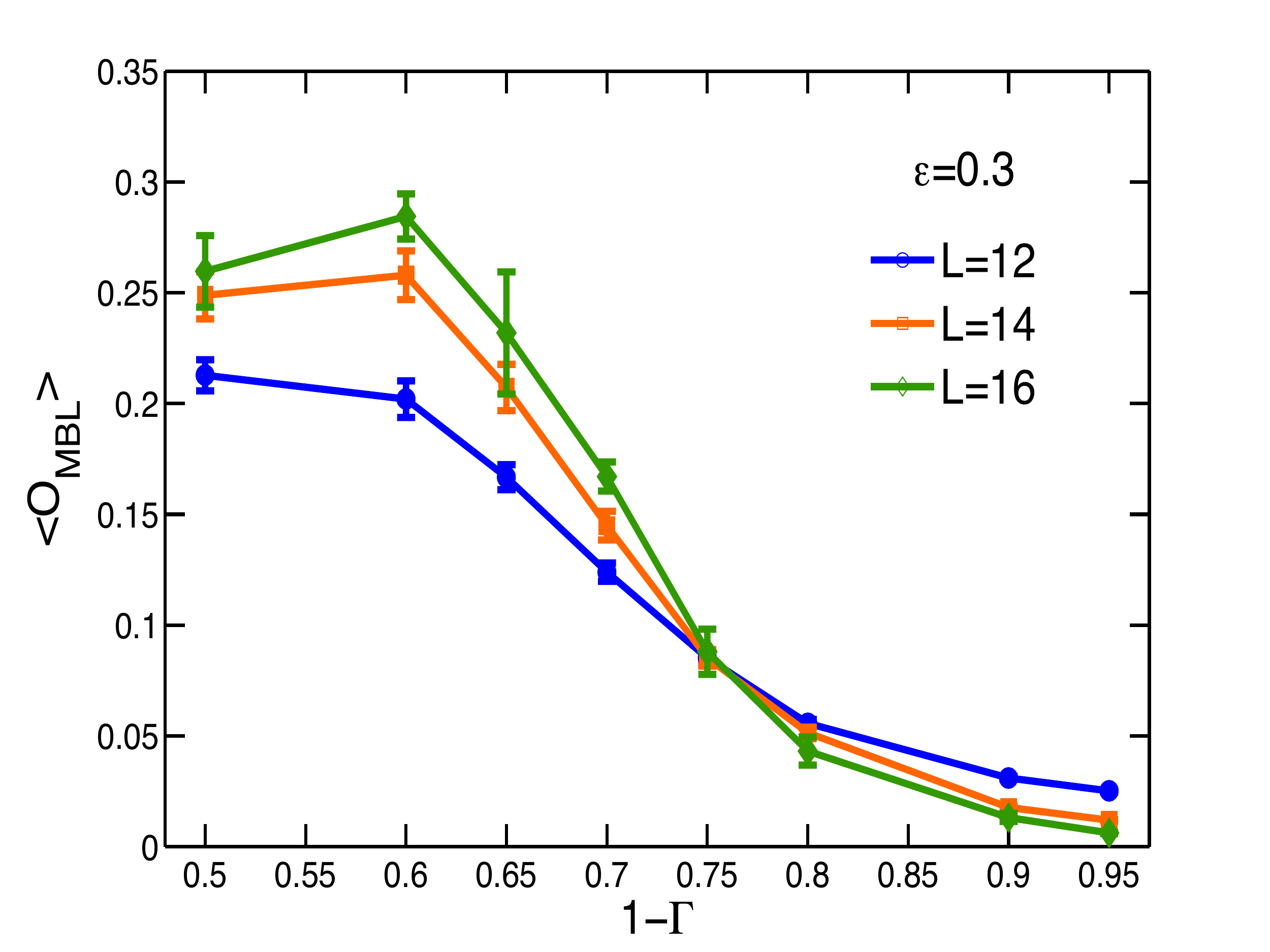}
\caption{The order parameter for QREM with target energy density
  $\epsilon=0.3$.}
\label{fig:QREM}
\end{figure*}

We also show the order parameter for the QREM at energy density
$\epsilon=0.3$, which is different from the one shown in the main
text. We clearly see from Fig. \ref{fig:QREM} that the curves for
different system sizes cross at around $\Gamma_c \approx 0.25$, which
is again in excellent agreement with previous known results.

\section{Randomness versus non-randomness: another toy model}

In this section, we view the emergence of the non-universal component
in the entanglement spectrum from a different perspective: the degree
of randomness in the wavefunctions. The physical intuition can be
understood as follows. A completely random state is supposed to yield
a Marchenko-Pastur distribution in its entanglement spectrum,
i.e. only the universal component exits. This implies that for states
whose entanglement spectra deviate from Marchenko-Pastur distribution,
they cannot be completely random. Therefore we construct another toy
model which captures this feature, by borrowing ideas from spin
glasses.

First let us start with a truly random (real) wavefunction that we
denote by $\Psi_{\rm REM}({\bm \sigma})={\rm sgn}(E_{\rm REM})$, where
${\rm sgn}(x)$ is the sign function and $E_{\rm REM}$ are identically
independently distributed with probability $P(E_{\rm
  REM})=\frac{1}{\sqrt{\pi L}}{\rm e}^{-E_{\rm REM}^2/L}$, with $L$
being the number of spins. The wavefunction $\Psi_{{\rm REM}}$ is, by
construction, random. The subscript REM is used to draw an analogy to
the Random Energy Model (REM) in spin glass systems\cite{16}.

\begin{figure}[!ht]
\centering
\includegraphics[width=0.45\textwidth]{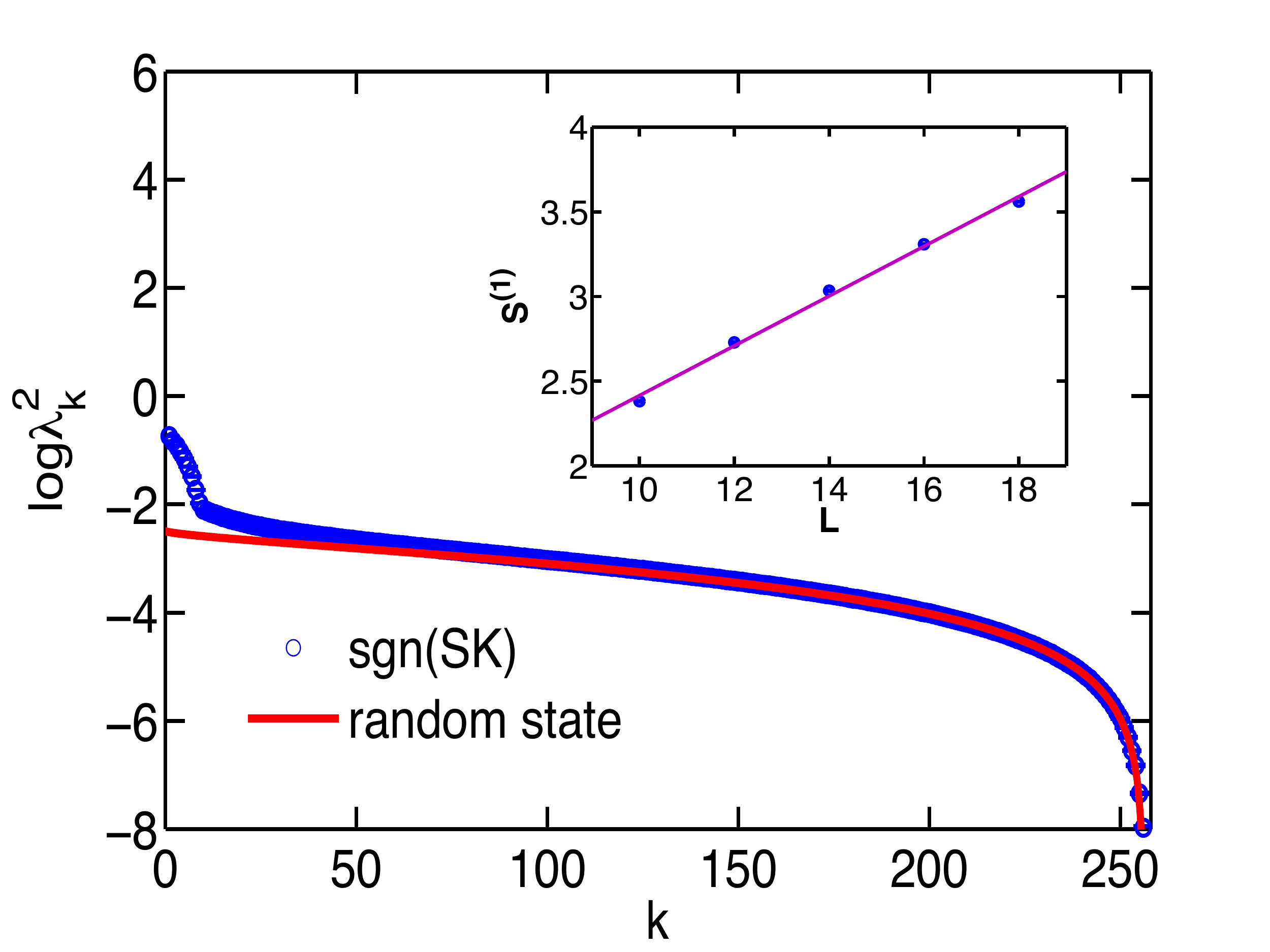}
\includegraphics[width=0.45\textwidth]{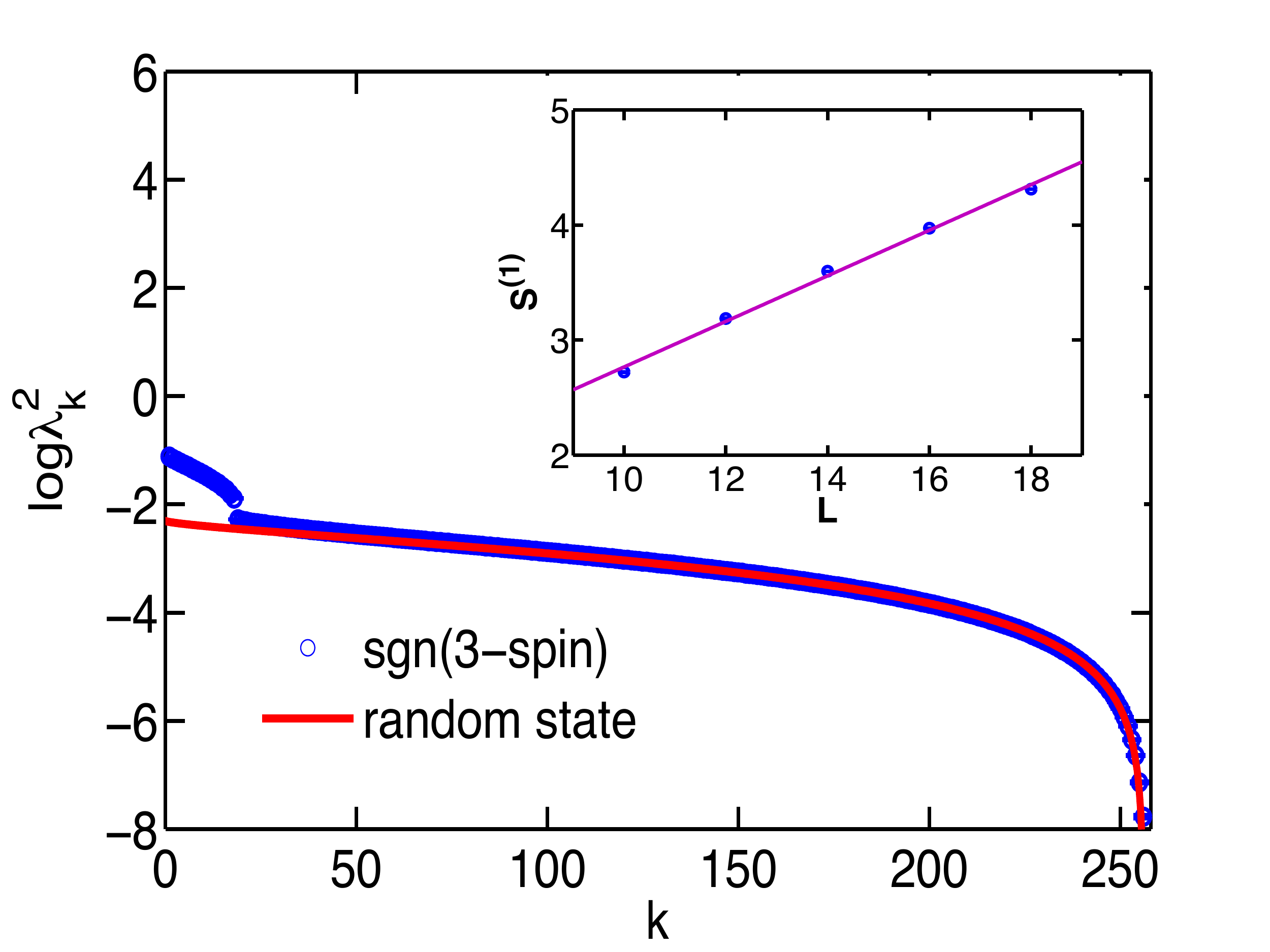}
\caption{(Color online) Entanglement spectrum of: $\Psi_{\rm SK}$
  (left) and $\Psi_{{\rm 3-spin}}({\bm \sigma})$ (right), for a system
  of size $L=16$, averaged over 500 realizations of disorder. The
  order parameter as defined in Eq. (3) is: $\langle {\mathcal
    O}\rangle = 0.3888\pm 0.0067$, with a threshold $k \approx 30$ for
  $\Psi_{\rm SK}$; and $\langle {\mathcal O}\rangle =0.4279\pm0.0060$,
  with a threshold $k\approx 24$ for $\Psi_{{\rm 3-spin}}({\bm
    \sigma})$. The insets show the volume-law scaling of the von
  Neumann entanglement entropy.}
\label{f5}
\end{figure}

Next, we note that the REM is a limiting case of a spin glass with
$p$-spin interactions as $p\rightarrow \infty$, which eliminates
correlations between configurations. So we can take a step back and
consider the following ``less random'' Sherrington-Kirkpatrick (SK)
spin glass model with infinite-range two-spin interactions \cite{SK},
and construct the following wavefunction:
\begin{equation}
 \Psi_{\rm SK}(\bm{\sigma}) = {\rm sgn}\left(E_{\rm SK}({\bm
   \sigma})\right)={\rm sgn}\left( \sum_{i<j} J_{ij}\;
 \sigma_i\,\sigma_j\right),
\end{equation}
where the $J_{ij}$ are drawn from uniform distribution in the interval
$[-1,1]$. The amplitudes computed from the SK-like model are obviously
not as random as in the REM-like one; there are only $L(L-1)/2$
independent random $J_{ij}$'s in the former as opposed to $2^L$
independent random amplitudes in the latter. Nevertheless, the
amplitudes of $\Psi_{\rm SK}$ do inherit some randomness from the
$J_{ij}$, and the energy distribution of the SK-like model also
follows accurately a Gaussian distribution for any {\it one} given
$\bm\sigma$, similarly to those in the REM-like model. But the
correlations between the amplitudes for {\it different} $\bm\sigma$
exist in the case of the SK-like model, and these correlations are
manifest in the entanglement spectrum computed from $\Psi_{\rm SK}$,
as shown in the left panel of Fig.~\ref{f5}.

The entanglement entropy follows a volume-law scaling (see inset of
Fig. \ref{f5}) and, again, we see the emergence of a two-component
structure in the spectrum. The universal part agrees with RMT. The
non-universal part is different from that found for the high energy
eigenstates of a disordered Heisenberg spin chain, reflecting its
non-universal, model dependent nature. Yet, this component is still
characterized by its fast decay rate. The toy model shows that non
randomness can be present in a generic quantum state when there are
correlations between components of the wavefunction. The entanglement
spectrum captures the non randomness and its structure can be well
described by a two-component picture.

\begin{figure}[!ht]
\centering
\includegraphics[width=0.7\textwidth]{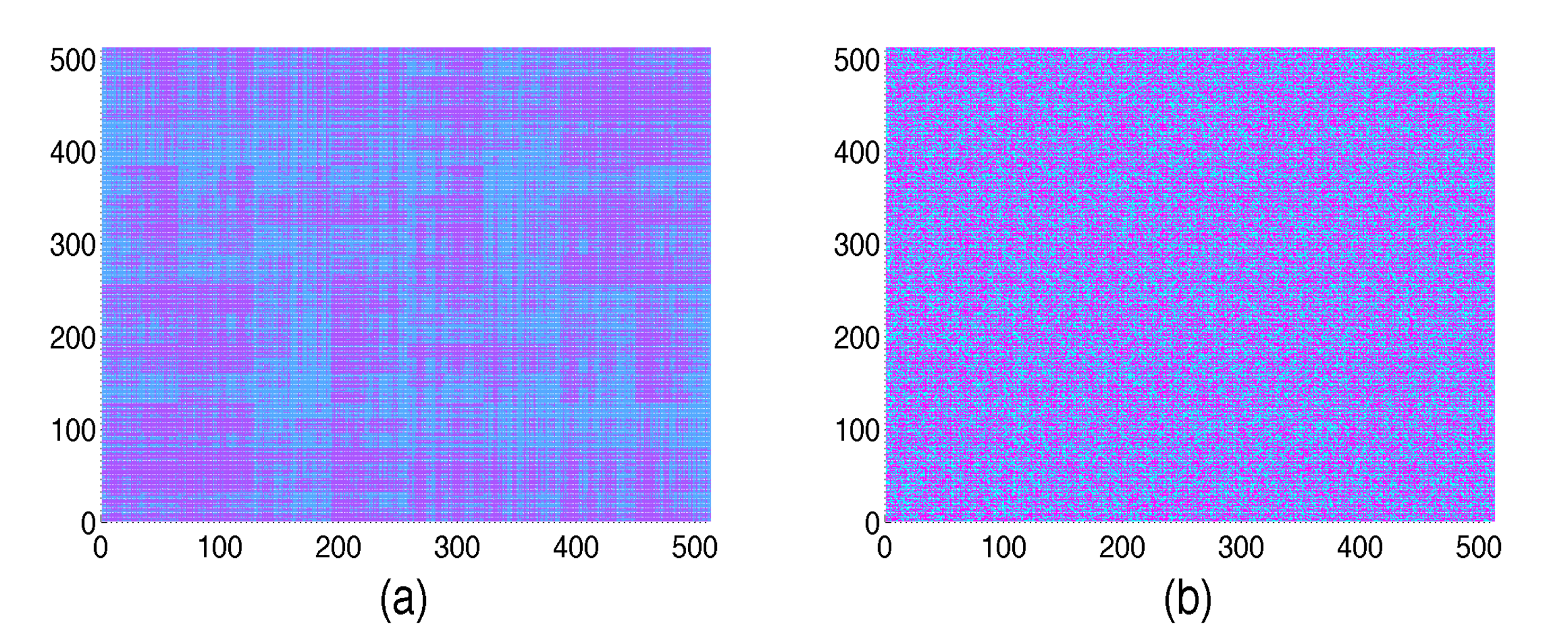}
\caption{(Color online) Color map of the matrix
  $\Psi(\bm{\sigma}_A,\bm{\sigma}_B)$ for typical realizations of the
  (a) SK and (b) REM wavefunctions ($L=18$).}
\label{fig:colormaps}
\end{figure}

Another interesting manifestation of the mixing between universal and
non-universal components in the SK wavefunction is revealed by
employing a color map. In Fig. \ref{fig:colormaps} we show the
amplitude of the wavefunction $\Psi_{\rm
  SK}(\bm{\sigma}_A,\bm{\sigma}_B)$ plotted in a $\bm{\sigma}_A \times
\bm{\sigma}_B$ grid, and compared it to the amplitude of a REM
wavefunction. The existence of a structure, similar to wefts in a
tapestry, is clearly visible for the $\Psi_{\rm SK}$ wavefunction, but
completely absent for the REM wavefunction.

One can further consider wavefunctions built with three-spin
interactions,
\begin{equation}
\Psi_{{\rm 3-spin}}({\bm \sigma}) = {\rm sgn}\left(\sum_{i<j<k}
J_{ijk}\, \sigma_i\, \sigma_j\, \sigma_k\right),
\end{equation}
where the $J_{ijk}$ are drawn from a uniform distribution in the
interval $[-1,1]$. As shown in the right panel of Fig. \ref{f5}, the
spectrum again shows a two-component structure, very similar to the
cases discussed in the main text.


\end{document}